\documentclass[11pt,bm,aps,nofootinbib,amsfonts,amssymb,preprintnumbers]{revtex4}

 \usepackage{here}

\usepackage[dvipdfmx]{graphicx}
\usepackage{amssymb,amsfonts,amsmath,bm,color,multirow,cases,empheq,hyperref}
\usepackage{mathrsfs}

\usepackage{enumerate}
\usepackage{ulem}
\usepackage{afterpage}

\newcommand{\cL}{{\cal L}}
\newcommand{\cF}{{\cal F}}
\newcommand{\cE}{{\cal E}}
\newcommand{\ex}{{\rm ex}}
\newcommand{\qstar}{{q_\star}}
\newcommand{\qdstar}{{q_{\star\star}}}

\hypersetup{%
 setpagesize=false,
 bookmarksnumbered=true,%
 bookmarksopen=true,%
 colorlinks=true,%
 linkcolor=blue,%
 citecolor=red}

\begin{document}




\title{Particle motions around regular black holes}

\author{Kenshin Isomura}
\email{sd17002@toyota-ti.ac.jp}
\author{Ryotaku Suzuki}
\email{sryotaku@toyota-ti.ac.jp}

\author{Shinya Tomizawa}
\email{tomizawa@toyota-ti.ac.jp}
\affiliation{Mathematical Physics Laboratory, Toyota Technological Institute\\
Hisakata 2-12-1, Nagoya 468-8511, Japan}
\date{\today}

\preprint{TTI-MATHPHYS-19}




\begin{abstract} 
We investigate the bound orbits of massive/massless, neutral particles and photons moving around regular black holes of Fan and Wang. 
For massive particles, we show the existence of stable/unstable circular orbits and
the charge dependence of the radius of the innermost stable circular orbit.
Remarkably, we find an unstable circular orbit of photons inside the event horizon.
For massless particles and photons, 
we show that both stable and unstable circular orbits can exist in a regular and horizonless spacetime with a slight overcharge.
Then, we also discuss the periapsis shift of massive neutral particles orbiting around the black hole,
and show that the charge gives a negative correction to the shift for black holes with small nonlinearity of electrodynamics.
\end{abstract}

\date{\today}
\maketitle



\section{Introduction}

The singularity theorems of Penrose and Hawking~\cite{Penrose:1964wq,Hawking:1970zqf} state that under the assumption of the presence of matter satisfying physically reasonable energy conditions, 
the  existence  of singularities is unavoidable in General Relativity (GR). 
However, it is widely believed that such singularities are simply nonphysical objects which are created by classical theories of gravity, and hence they will be resolved if we can obtain complete quantum gravity in our future. 
Bardeen~\cite{Bardeen:1968} proposed the first model of asymptotically flat, static and spherically symmetric black holes (BHs) with a regular center. Such a kind of BH is called a regular black hole (RBH) or a non-singular BH. 
At first, this BH model was not obtained as an exact solution of the Einstein equation, but thereafter, Ay\'on-Beato and Garc\'ia~\cite{Ayon-Beato:2000mjt} showed that the Bardeen model can be seen as a solution of the Einstein equation  coupled with a physical source of a magnetic monopole in nonlinear electrodynamics (NED). 
They also found another type of RBH solution which describes a Reissner-Nordstr\"om type spacetime to the Einstein-NED equation~\cite{Ayon-Beato:1998hmi}. 
Subsequently,  other RBH models were proposed. 
For instance, Dymnikova~\cite{Dymnikova:1992ux} proposed a different type of RBH, which coincides with the Schwarzschild spacetime near infinity and behaves like the de Sitter spacetime near a center.
Hayward proposed a static and spherically symmetric RBH model to resolve the BH information-loss paradox~\cite{Hayward:2005gi}. 
Furthermore, Fan and Wang~\cite{Fan:2016hvf} found a wide class of asymptotically flat, static and spherically symmetric RBH solution in NED, which generalize the Bardeen BH~\cite{Bardeen:1968} and the Hayward BH~\cite{Hayward:2005gi}. 
Other than these RBHs, numerous types of models and solutions have been proposed so far.
The readers can find useful reviews in refs.~\cite{Ansoldi:2008jw,Lemos:2011dq,Maeda:2021jdc}.

\medskip

The observation of particle motion around BHs is useful to test GR and alternative theories of gravity since it enables one to give the constraints on the parameters of the spin and charge of BH. 
So far, many researchers have also studied particle motion around RBHs. 
Ref.~\cite{Zhou:2011aa,Garcia:2013zud, Stuchlik:2014qja} investigated circular geodesics in the Bardeen BH and Ay\'on-Beato-Garc\'ia BH. 
The gravitational lens of the Bardeen BH was discussed in~\cite{Eiroa:2010wm}.
Moreover, Ref.~\cite{Stuchlik:2019uvf} studied the photon orbits around the Bardeen BH, and determined the BH shadow. 
Ref.~\cite{Gao:2020wjz} studied the periapsis shifts of bound orbits of massive particles moving around Bardeen BHs.
Ref.~\cite{Rayimbaev:2020hjs} investigated the particle motion around a special class of the Fan-Wang BH (FWBH) with Maxwell weak-field limit.
Ref.~\cite{Carballo-Rubio:2022nuj} studied the particle motion around the Hayward BH. 
A remarkable aspect in non-linear electrodynamics is that photons do not propagate along the null geodesics of the spacetime geometry but rather of an {\it effective geometry}~\cite{Novello:1999pg,Novello:2000km,Stuchlik:2019uvf,Rayimbaev:2020hjs}. 
The propagation of photons has been studied for the Ay\'on-Beato-Garc\'ia spacetime in Ref.~\cite{Novello:2000km}, for the Bardeen spacetime in Ref.~\cite{Stuchlik:2019uvf} and for the Hayward spacetime in Ref.~\cite{Toshmatov:2019gxg}. The photon orbits are also studied in  rotating versions of several RBHs of NED~\cite{Kumar:2020ltt}.
In this article, we aim to study the motion of massive/massless neutral particles and photons in the FWBH spacetime and derive the general properties of RBHs spacetime because the RBH covers the Bardeen BH and Hayward BH spacetimes as special cases.

\medskip
The rest of the paper is devoted to analyze the particle motion around the FWBHs. 
In the next section, we review the FWBHs as solutions in NED,  then present the metric and the gauge potential with a magnetic monopole or an electric charge, and further give the conditions for the existence of horizon.
In Sec.~\ref{sec:massive}, we discuss the stability of circular orbits for massive particles around the FWBHs, where the particle motion can be reduced to a one-dimensional potential problem.
In Sec.~\ref{sec:massless},  we similarly consider the motion for massless particles, whose potential can be obtained as the divergence limit of an angular momentum. 
However, the massless orbits do not correspond to photon orbits in the Einstein gravity coupled with NED.
Hence, in Sec.~\ref{sec:photon}, we separately analyze photon orbits around the FWBHs. 
In Sec.~\ref{sec:shift}, we compute the periapsis shift of the orbits for massive particles  in the weak-field limit. 
In Sec.~\ref{sec:conclusion}, we summarize our results and discuss possible generalization.



\section{Review of regular black holes}\label{sec:solution}
Here we review the RBHs of general Fan-Wang class~\cite{Fan:2016hvf}, which are given as the solution with NED whose action is given by
\begin{align}
 S = \frac{1}{16 \pi G} \int dx^4 \sqrt{-g} (R - {\cal L}({\cal F})),\quad {\cal F} := F_{\alpha\beta} F^{\alpha\beta}.
\end{align}
The field equations derived from the Lagrangian density $\cL(\cF)$ of NED admit electrically charged solutions or magnetically charged solutions, where in particular, $\cL(\cF)$ for the  latter case can  be written as   
\begin{align}
 {\cal L}({\cal F})=\frac{4\mu}{\alpha}\frac{(\alpha{\cal F})^{\frac{\nu+3}{4}}}{  \left(1+(\alpha{\cal F})^{\frac{\nu}{4}} \right)^{\frac{\mu+\nu}{\nu}}  },\label{eq:LF-magnetic}
\end{align}
where $\mu,\nu$ and $\alpha$ are free parameters of the theory.
This theory reproduces the usual Maxwell theory at the weak field limit ${\cal F}\to 0$ only if $\nu=1$, which is the main subject of ref.~\cite{Rayimbaev:2020hjs}.
The asymptotically flat, static and spherically symmetric BH solution with a magnetic monopole is given by~\cite{Fan:2016hvf}
\begin{eqnarray}
ds^2&=&-f(r)dt^2+\frac{dr^2}{f(r)}+r^2(d\theta^2+\sin^2\theta d\phi^2), \quad f(r)= 1-\frac{2M-2q^3\alpha^{-1}}{r}-\frac{2 \alpha^{-1}q^3 r^{\mu-1}}{(r^\nu+q^\nu)^{\frac{\mu}{\nu}}}, \label{eq:metric0}
\end{eqnarray}
and
\begin{align}
 A= \frac{q^2}{\sqrt{2\alpha}} \cos\theta d\phi,\quad \cF = \frac{q^4}{\alpha r^4},\label{eq:F-magnetic}
\end{align}
where $M$ is the ADM mass of the spacetime\footnote{See also a comment~\cite{Toshmatov:2018cks} on the definition of the mass.}.  The magnetic charge is given by
\begin{align}
&  Q_m := \frac{1}{4\pi} \int F = \frac{q^2}{\sqrt{2\alpha}}.
\end{align}
On the other hand,  the gauge potential for an electrically charged BH solution with the same metric~(\ref{eq:metric0}) can be written as
\begin{align}
 A = \frac{q^2}{2\alpha}\left(\frac{r^\mu(3r^\nu-(\mu-3)q^\nu)}{(r^\nu+q^\nu)^{\frac{\mu+\nu}{\nu}}}-3\right)dt,\label{eq:A-electric}
\end{align}
where $q$ is now related to the electric charge defined by
\begin{align}
 Q_e := \frac{1}{4\pi} \int \cL_{\cF} \star F = \frac{q^2}{\sqrt{2\alpha}}.
\end{align}
The scalar $\cF$ is given by
\begin{align}
\cF =  -\frac{\mu ^2 q^{2 \nu +4} r^{2 \mu -2} \left(q^{\nu }+r^{\nu
   }\right)^{-\frac{2 \mu }{\nu }-4} \left((\mu -3) q^{\nu }-(\nu +3)
   r^{\nu }\right)^2}{2 \alpha ^2}.\label{eq:L-electric}
\end{align}
Unlike the magnetic solution, one cannot obtain the explicit form of the NED Lagrangian, but only its on-shell value for the solution as
\begin{align}
\cL(\cF)=\frac{ 2  q^{3+\nu} r^{\mu-3}}{\alpha} \frac{(\mu-1)q^\nu-(\nu+1)r^\nu}{(q^\nu+r^\nu)^{2+\frac{\mu}{\nu}}}.
\end{align}
In order to eliminate the singularity at the center $r=0$, one must set
\begin{align}
 M = \alpha^{-1} q^3\label{eq:Mtoalpha},
\end{align}
in which the metric function $f(r)$ can be written as
\begin{eqnarray}
f(r) = 1 - \frac{2M r^{\mu-1}}{(r^\nu+q^\nu)^\frac{\mu}{\nu}}. \label{eq:metric}
\end{eqnarray}
This metric admits known RBH spacetimes for specific parameter choices
\begin{itemize}
\item $(\mu,\nu)=(3,2)$ : Bardeen BHs~\cite{Bardeen:1968}
\item $(\mu,\nu)=(3,3)$ : Hayward BHs~\cite{Hayward:2005gi}
\end{itemize}
Although the mass and charge are constrained by eq.~(\ref{eq:Mtoalpha}), 
we can treat them as independent parameters by adjusting the parameter $\alpha$.

In the following, we use the gravitational radius $r_g$ instead of the mass $M$,
\begin{align}
 r_g := 2M.
\end{align}

The equation $f(r)=0$ has two roots $r_-$ and $r_+$ ($r_-<r_+$) for $r>0$, which corresponds to an inner horizon and an outer horizon. 
Let us consider the extreme condition ($r_+=r_-$)  given by $f(r)=0$ and $f'(r)=0$, which can be solved as
\begin{eqnarray}
r&=&(\mu-1)^{\frac{1}{\nu}}q=r_g \frac{(\mu-1)^{\frac{\mu}{\nu}} } 
{\mu^{\frac{\mu}{\nu}}}:=r_{h}^\ex
 ,
\\
q&=&r_g\frac{(\mu-1)^{\frac{\mu-1}{\nu}} } 
{\mu^{\frac{\mu}{\nu}}}:=q^\ex.
\end{eqnarray}
Therefore, we have four distinct cases depending on the charge:
\begin{enumerate}
\item a black hole with a single horizon (Schwarzschild  BH) : $q=0$
\item a black hole with two horizons : $0<q<q^\ex$
\item a degenerate horizon : $q=q^\ex$
\item an overcharged  but regular horizonless spacetime: $q>q^\ex$
\end{enumerate}

It is straightforward to show $r_h^\ex$ monotonically increases with respect to both parameters $\mu$ and $\nu$, while 
$q^\ex$ monotonically decreases with respect to $\mu$ and increases with respect to $\nu$. At large values of $\mu$ and $\nu$, $r_h^\ex$ and $q^\ex$ approaches to the following limits,
\begin{align}
&  r^\ex_h/r_g \to \left\{\begin{array}{cc} e^{-1/\nu}& ( \mu \to \infty) \\ 1&(\nu\to\infty)  \end{array}\right. ,\\
&   q^\ex/r_g \to \left\{\begin{array}{cc} (\mu e)^{-1/\nu}& ( \mu \to \infty) \\ 1&(\nu\to\infty)  \end{array}\right. .\label{eq:qex-limits}
\end{align}
In particular, $q^\ex$ tends to be zero for large $\mu$.
For the later use, we also show
\begin{align}
\frac{\partial (\mu^{1/\nu} q^\ex(\mu,\nu))}{\partial \mu} < 0
\end{align}
for fixed $\nu$. Together with eq.~(\ref{eq:qex-limits}), this determines the range of $q^\ex$ for a given $\nu$ as 
\begin{align}
e^{-1/\nu} \leq   \mu^{1/\nu} q^\ex/r_g \leq \left(2/3\right)^{2/\nu}\quad (\mu \geq 3).  
\label{eq:qex-range}
\end{align}

\section{Circular orbits of massive particles}\label{sec:massive}
The particle motion in the FW spacetimes~(\ref{eq:metric}) is determined by the Lagrangian
\begin{align}
 {\cal L}_p = \frac{1}{2} g_{\mu\nu} \dot{x}^\mu \dot{x}^\nu 
\end{align}
where the dot denotes the derivative with respect to an affine parameter. The motion should satisfy the constraint $g_{\mu\nu} \dot{x}^\mu \dot{x}^\nu =-\kappa$, in which $\kappa=1$ is for massive and $\kappa=0$ for massless particles. The spherical symmetry of the spacetime allows us to assume the movement takes place in the equatorial surface $\theta = \pi/2$ without loss of generality.
Since the metric is independent of $t$ and $\phi$, their conjugates give two constants of motion, the energy and angular momentum
of the particle
\begin{align}
 \cE := -\frac{\partial {\cal L}_p}{\partial \dot{t}} =  f \dot{t} ,\quad L := \frac{\partial {\cal L}_p}{\partial \dot{\phi}}=r^2 \dot{\phi}.
 \label{eq:geodesic-t-phi}
\end{align}
Then, the equation of motion can be written as a one-dimensional motion in the effective potential
\begin{eqnarray}
\dot{r}^2  + U(r) = \cE^2 ,\quad  U(r) :=f(r) \left(\kappa+\frac{L^2}{r^2}\right). \label{eq:geodesic-r-U}
\end{eqnarray}
Without loss of generality, we may assume $L\geq 0$.
In this section, we consider the motion of massive particles by setting $\kappa=1$.
In particular, we focus on the circular orbit which corresponds to stationary points of $U$,
\begin{align}
U = \cE^2,\quad U_{,r} = 0.\label{eq:cond-circular}
\end{align}
By fixing the charge parameter $q$, the angular momentum $L$ and energy $\cE$ are given by functions of the orbit radius $r_c$ as 
\begin{align}
L^2 (r_c)= \frac{r_g r_c^{\mu +2} \left(r_c^{\nu }-(\mu -1) q^{\nu }\right)}{q^{\nu } \left(2r_c \left(q^{\nu
   }+r_c^{\nu }\right)^{\mu /\nu }+(\mu -3) r_g r_c^{\mu }\right)+r_c^{\nu } \left(2r_c \left(q^{\nu
   }+r_c^{\nu }\right)^{\mu /\nu }-3 r_g r_c^{\mu }\right)},\label{eq:def-L2}
\end{align}
and
\begin{align}
{\cal E}^2(r_c) = \frac{2\left(q^{\nu }+r_c^{\nu }\right)^{1-\frac{\mu }{\nu }} \left(r_c \left(q^{\nu }+r_c^{\nu  }\right)^{\mu /\nu }-r_g r_c^{\mu }\right)^2}{r_c \left(q^{\nu } \left(2r_c \left(q^{\nu
   }+r_c^{\nu }\right)^{\mu /\nu }+(\mu -3) r_g r_c^{\mu }\right)+r_c^{\nu } \left(2r_c \left(q^{\nu
   }+r_c^{\nu }\right)^{\mu /\nu }-3r_g r_c^{\mu }\right)\right)}.
\end{align}
Note that physical circular orbits should also satisfy $L^2 \geq 0$ and $\cE^2 \geq 0$.
The signature of $dL/dr_c$ coincides with that of $U_{,rr}(r_c,L(r_c))$ as
\begin{align}
U_{,rr}   = - U_{,rL}  \frac{dL}{dr_c} ,\quad -U_{,rL} = \frac{4Lf}{r(r^2+L^2)}>0,
\end{align}
and hence the stable circular orbits are given equivalently by $U_{,rr}>0$ or $dL/dr_c>0$. Similarly, one can show 
\begin{align}
\frac{d \cE^2}{dr_c} = \frac{dL}{dr_c} \frac{2 L f}{r^2},
\end{align}
which means $d\cE^2/dr_c$ also has the same signature as $U_{,rr}$.

As in the Schwarzschild case, circular orbits 
cannot exist sufficiently close to the horizon.
The inner edge of the orbital area is characterized by the innermost stable circular orbit (ISCO), which is given by
\begin{align}
 U = \cE^2,\quad U_{,r} = 0,\quad U_{,rr}=0. \label{eq:ISCO-condition-1}
\end{align}
With the discussion above, the condition $U_{,rr}=0$ is equivalent to $dL/dr_c=0$.

However, we find that this only gives the ISCO as far as the horizon exists, i.e. $q\leq q^\ex$.
For the horizonless case ($q>q^\ex$), the ISCO is given by the orbit with zero angular momentum
\begin{align}
 U = \cE^2,\quad U_{,r}=0, \quad L=0. 
\end{align}
This ``orbit" corresponds to a particle in the static equilibrium between the repulsive force by the inner de Sitter-like core and outer attraction.
The radius for $L=0$ is explicitly given by
\begin{align}
 r = r_0(q) := (\mu-1)^{1/\nu} q.\label{eq:ISCO-condition-2}
\end{align}

\begin{figure}[H]
\begin{center}
\includegraphics[width=7.6cm]{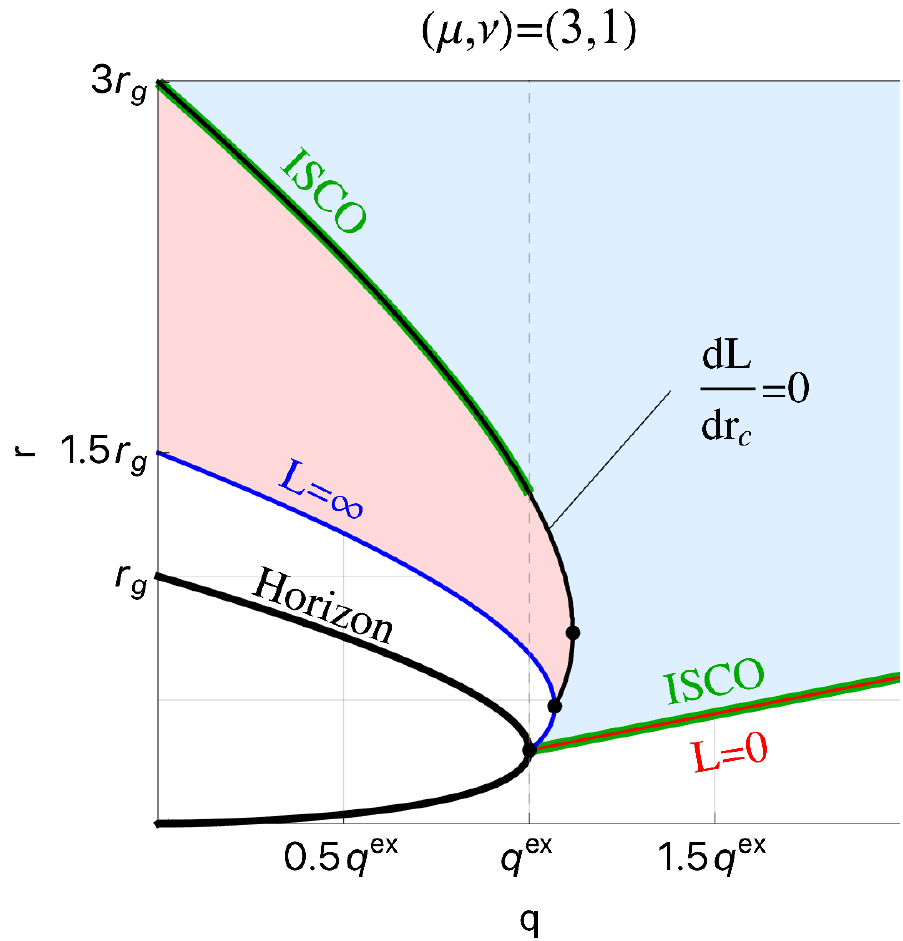}\hspace{4mm}
\includegraphics[width=4.5cm]{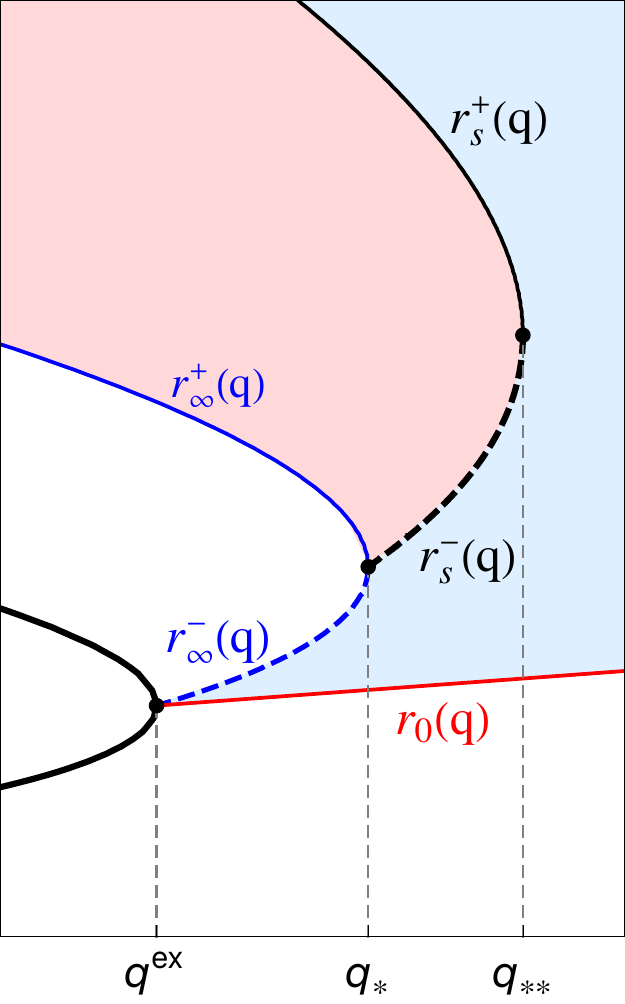}
\caption{Existence and stability of circular orbits in the Fan-Wang spacetime, plotted in $(q,r)$-space for $(\mu,\nu)=(3,1)$. The regions colored by light blue has stable circular orbits, and light red only has unstable ones. Circular orbits do not exist in the white region. The right figure is a closeup around $q=q^\ex$. As we mention later, the blue curve corresponds to massless orbits which splits to two branches $r^\pm_{\infty}$.\label{fig:qrplot31}}
\end{center}
\end{figure}

In Fig.~\ref{fig:qrplot31}, we plot the positions of stable and unstable circular orbits for a given charge.
Although we only show the result for $(\mu,\nu)=(3,1)$, other cases are qualitatively the same.
One can see that the ISCO is given by eq.~(\ref{eq:ISCO-condition-1}) for $0\leq q \leq q^\ex$, and by eq.~(\ref{eq:ISCO-condition-2}) for $q>q^\ex$.
For $q>q^\ex$, although the curve by eq.~(\ref{eq:ISCO-condition-1}) is not the ISCO, it still gives two branches of solution $r=r_s^\pm(q)$ for a slightly overcharged case, where we have an unstable orbits between the two $r_s^- < r < r_s^+$. This unstable region disappears for $q\geq \qdstar$, where the threshold value is given by $r_{\star\star}:=r_s^+(\qdstar)=r_s^-(\qdstar)$.

Since both $\cE$ and $L$ must be real, circular orbits do not exists if the solution of eq.~(\ref{eq:cond-circular}) gives either $L^2<0$ or $\cE^2<0$. The border of the existence region is given by both $L=0$ and $L=\infty$. The $L=\infty$ curve corresponds to the circular orbits for massless particles which is discussed later. This massless curve also has two branches $r=r_\infty^\pm(q)$ bifurcating at $(\qstar,r_\star)$ for a slightly overcharged case, which is already observed in the previous study of $\nu=1$ cases~\cite{Stuchlik2019}. One should note that all these characteristics, in fact, also appear in the Reissner-Nordstr\"om spacetime~\cite{Pugliese:2010ps}.

\begin{figure}[H]
\begin{center}
\includegraphics[width=8cm]{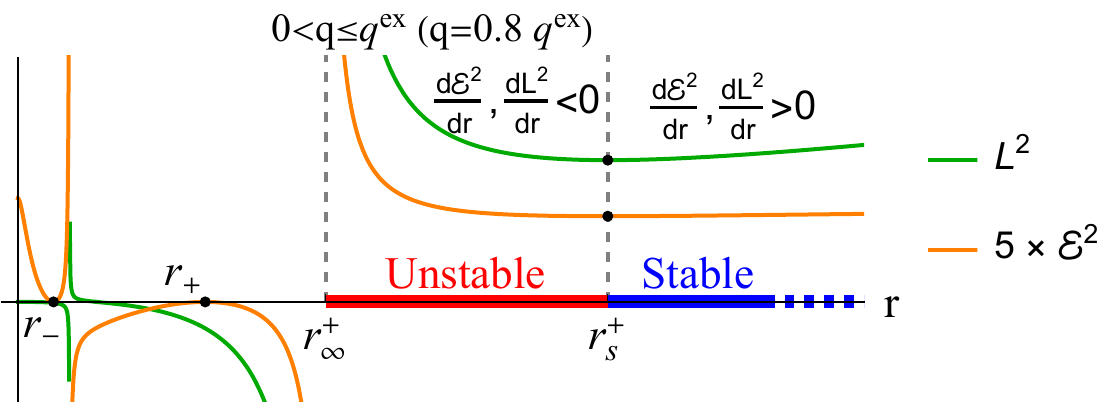}\hspace{0.2cm}
\includegraphics[width=8cm]{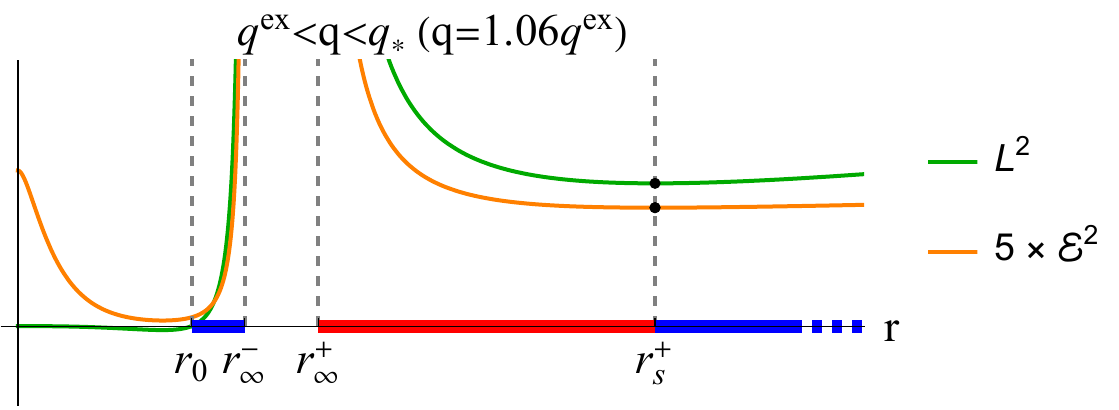}\\
\includegraphics[width=8cm]{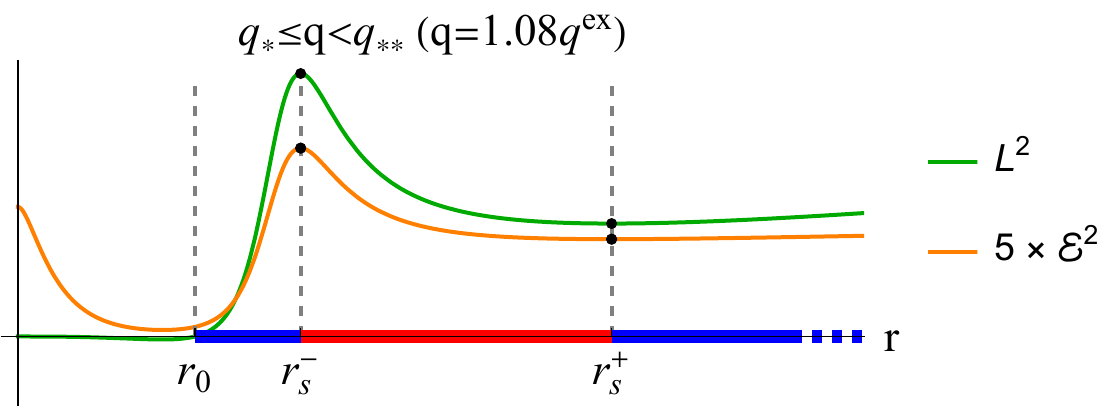}\hspace{0.2cm}
\includegraphics[width=8cm]{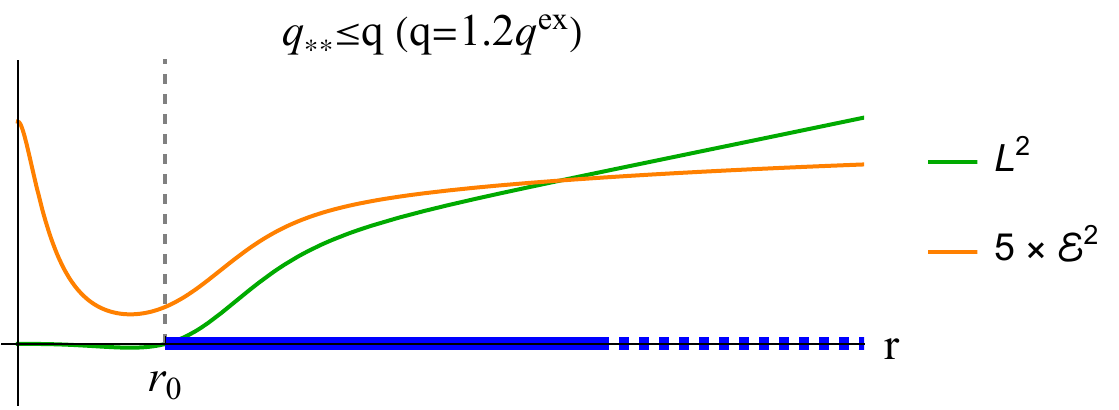}
\caption{Stability of circular orbits are compared with gradients of $L^2$ and $\cE^2$ for each $q$. Orbits only exist if $\cE^2>0$ and $L^2>0$. \label{fig:orbit_cases}}
\end{center}
\end{figure}

As shown in Fig.~\ref{fig:orbit_cases}, the branching points for $r^\pm_\infty(q)$ and $r^\pm_s(q)$, which we denote $\qstar$ and $\qdstar$ respectively, gives threshold values of $q$ that causes qualitative changes on the circular orbits.

\begin{figure}[H]
\begin{center}
\includegraphics[width=16cm]{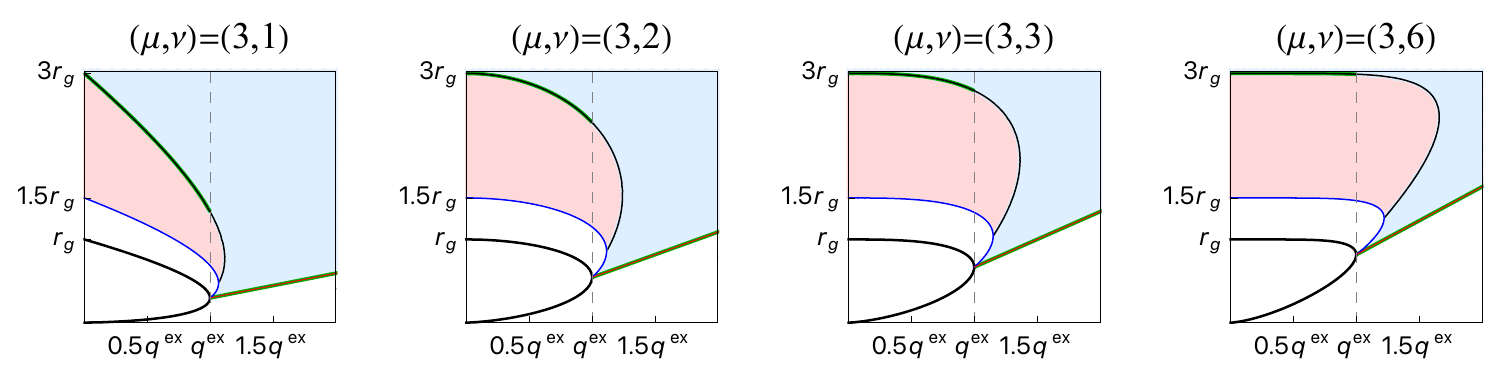}
\includegraphics[width=16cm]{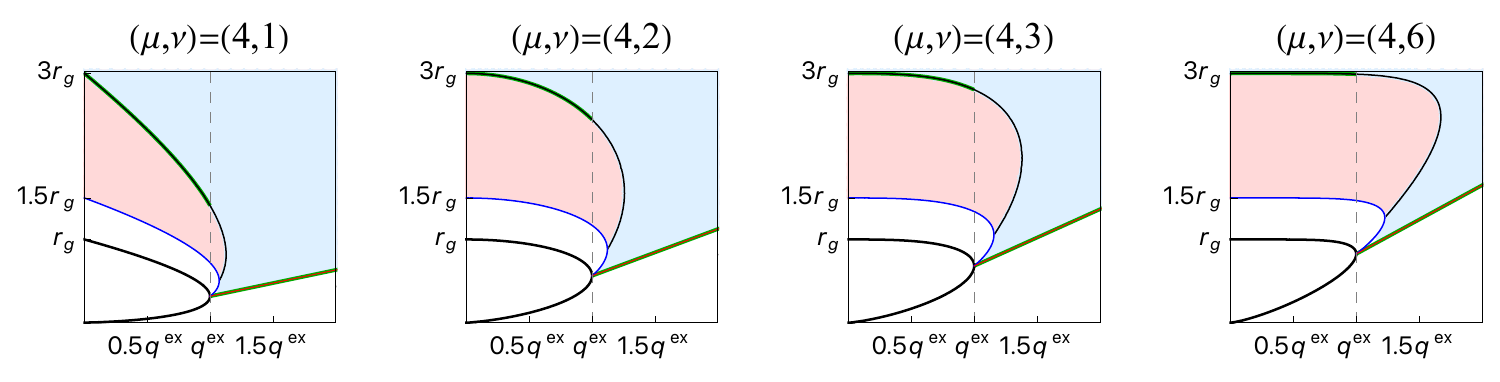}
\caption{Circular orbits for $\mu=3,4$ and $\nu=1,2,3$ and $6$. Curves in the plots correspond to those in fig.~\ref{fig:qrplot31}.\label{fig:qrplots3nu}}
\end{center}
\end{figure}

\begin{figure}[H]
\begin{center}
\includegraphics[width=15cm]{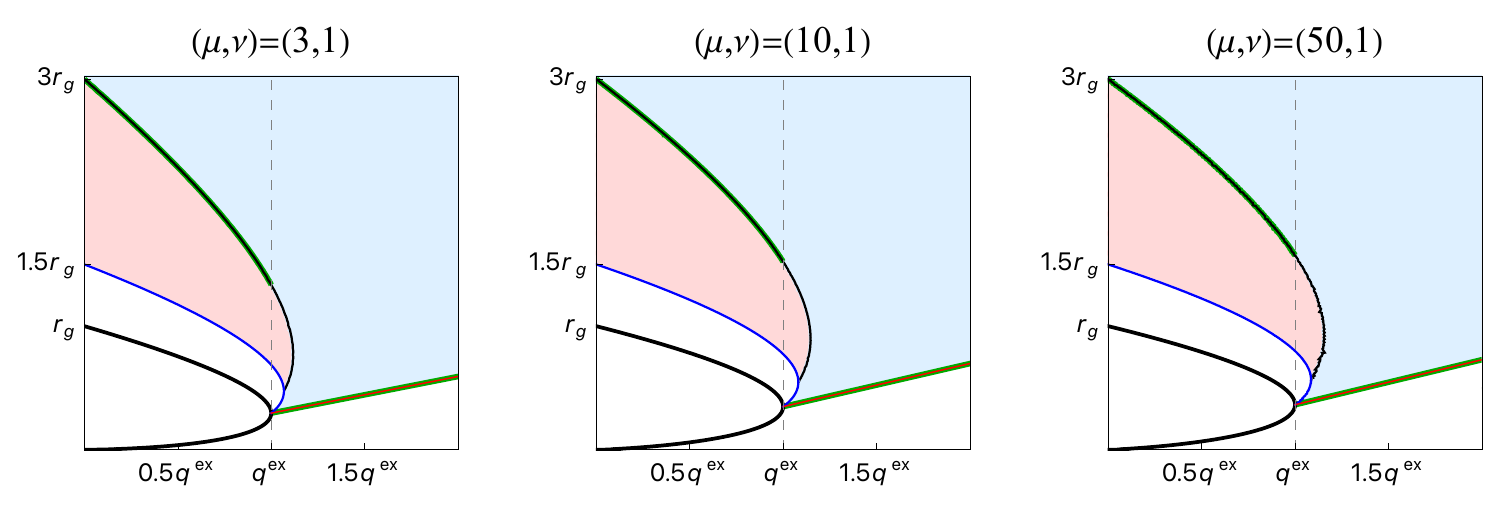}\\
\includegraphics[width=15cm]{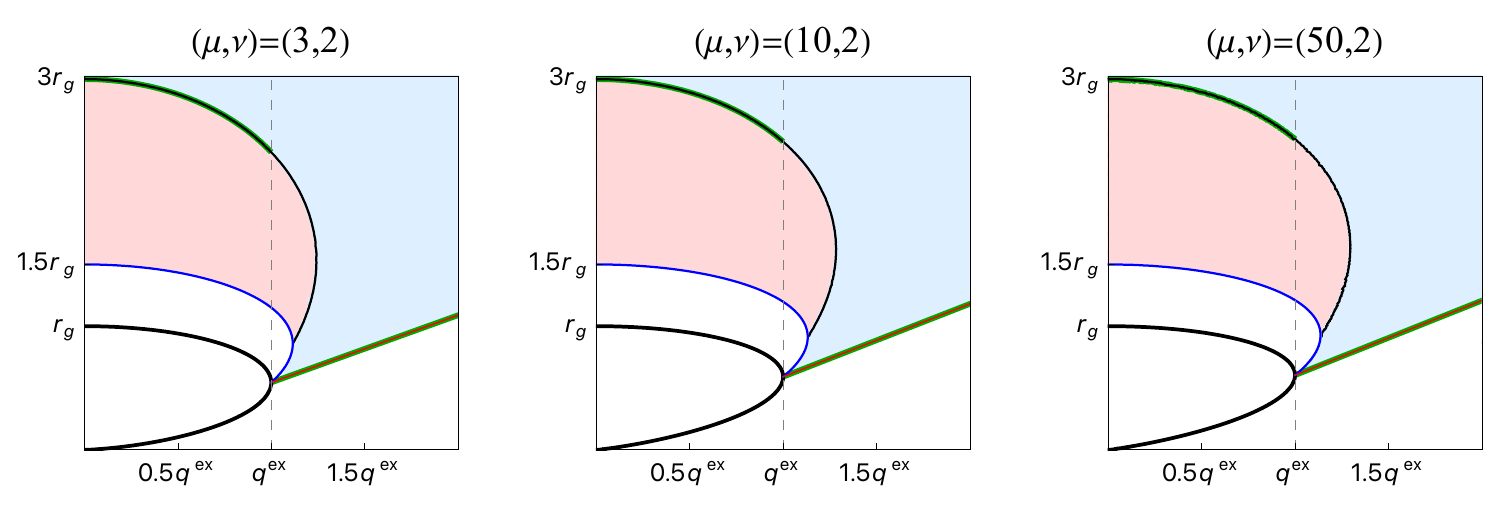}
\caption{Circular orbits for $\nu=1,2$ with $\mu=3,10,50$. Curves in the plots correspond to those in fig.~\ref{fig:qrplot31}.\label{fig:qrplotsmu1}}
\end{center}
\end{figure}

We also study the effect of parameters $\mu$ and $\nu$ on circular orbits.
For $\nu=1$, the $\mu$-dependence was studied in the previous work~\cite{Rayimbaev:2020hjs}.
In Fig.~\ref{fig:qrplots3nu}, the $\nu$-dependence of circular orbits are shown for $\mu=3,4$. 
In each cases, as $\nu$ grows, the appearance of circular orbits approach to a certain shape.
Especially, the ISCO curve and massless curve approaches to almost straight lines for $q\leq q^\ex$.
In Fig.~\ref{fig:qrplotsmu1}, the $\mu$-dependence is studied as well. Unlike the $\nu$-dependence, we find that the appearance of circular orbits is quite insensitive to the change in $\mu$.

\begin{figure}[H]
\begin{center}
\includegraphics[width=8cm]{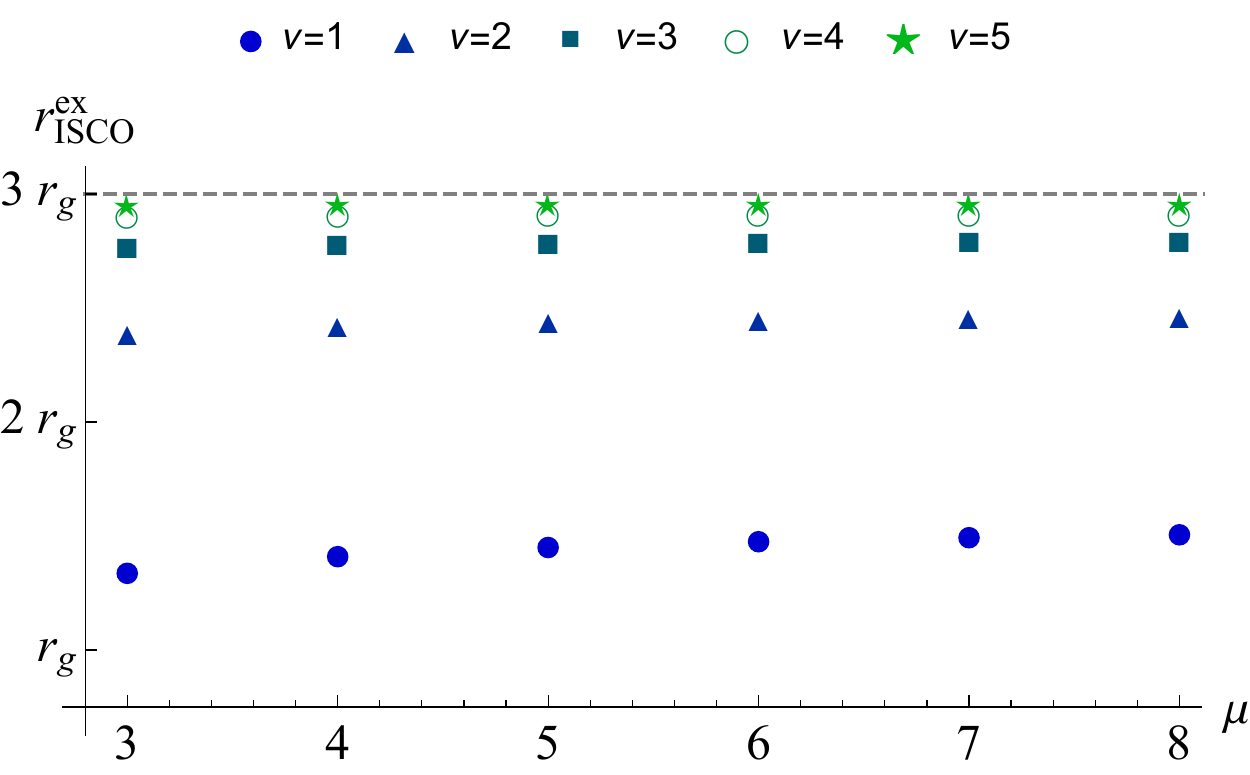}
\includegraphics[width=8cm]{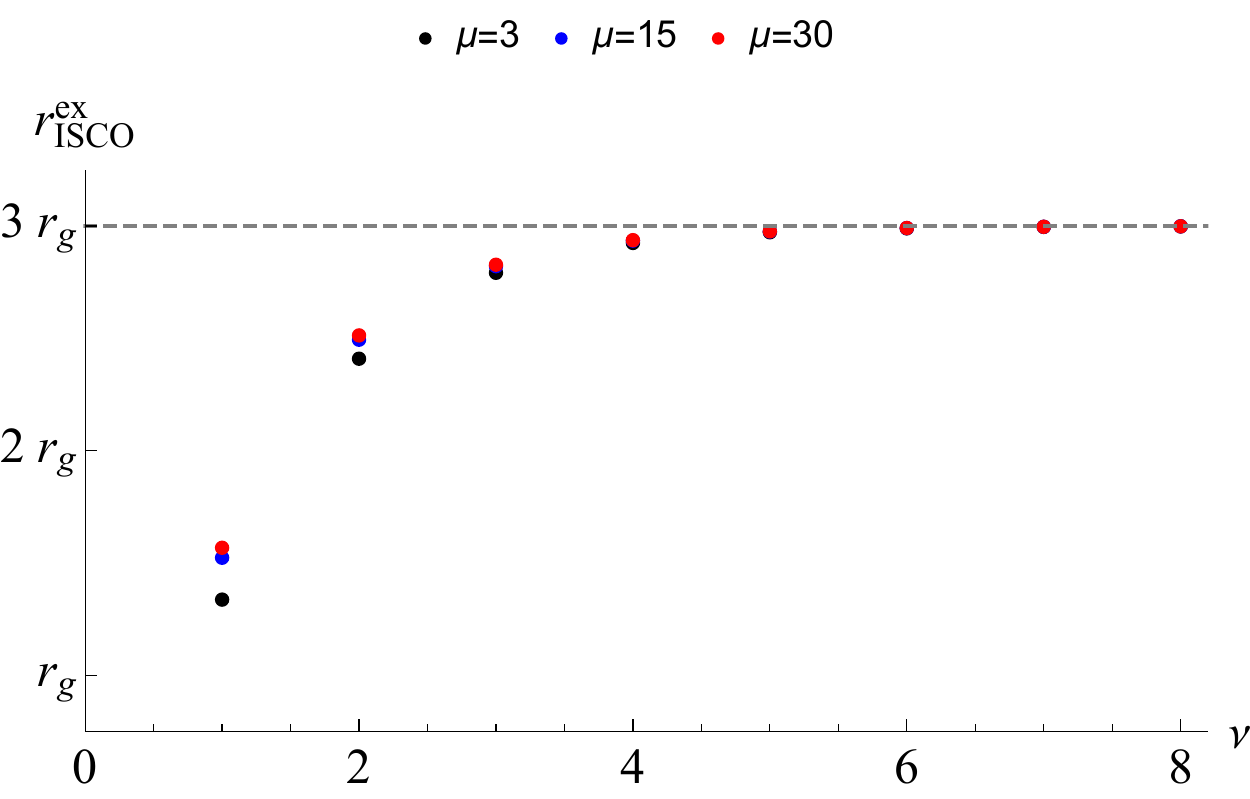}
\caption{$\mu$ and $\nu$-dependence of $r_{\rm ISCO}^\ex$. $\nu$ is fixed in the left panel and $\mu$ is fixed in the right panel. \label{fig:riscoex}}
\end{center}
\end{figure}

The parameter dependence of the ISCO radii around the black hole can be roughly estimated by the value at $q=q^\ex$, say $r^\ex_{\rm ISCO}=r^+_s(q^\ex)$, as it gives the minimum value of $r_{\rm ISCO}$ for $q \leq q^\ex$. Fig.~\ref{fig:riscoex} shows $r^\ex_{\rm ISCO}$ is a monotonically increasing function of $\mu$ and $\nu$. 

\section{Circular orbits of massless particles}\label{sec:massless}
The effective potential for the massless particle is given by setting $\kappa=0$ in eq.~(\ref{eq:geodesic-r-U}), 
\begin{eqnarray}
V:= \frac{f}{r^2}.
\end{eqnarray}
The radius of the circular orbit for massless particles is determined by 
\begin{eqnarray}
V_{,r}(r_c,q)
=0,
\end{eqnarray}
which is explicitly written as
\begin{align}
 3 r_c^\nu - (\mu-3) q^\nu = 2 r_g^{-1} r_c^{1-\mu} ( r_c^\nu+q^\nu) ^\frac{\mu+\nu}{\nu},\label{eq:massless-Vr-expr}
\end{align}
This is equivalent to the condition~(\ref{eq:cond-circular}) with $L=\infty$, and hence the position of circular orbits are given as $r_\infty^\pm(q)$ in Fig.~\ref{fig:qrplot31}.

Differentiating the orbit $q=q(r_c)$, we obtain
\begin{eqnarray}
V_{,rr}+V_{,rq}\frac{dq}{dr_c}=0.
\end{eqnarray}
This leads to
\begin{eqnarray}
\frac{dq}{dr_c}=\frac{(r_c^\nu+q^\nu)^{\frac{\mu+2\nu}{\nu}}}{\mu r_g^2 q^{\nu-1}(\nu r_g r_c^{\mu+\nu-4}+2r_c^{-3}(r_c^\nu+q^\nu)^\frac{\mu+\nu}{\nu})}V_{,rr},
\end{eqnarray}
where we have used eq.~(\ref{eq:massless-Vr-expr}). From the positivity of the factor before $V_{,rr}$, one can show that $dq/dr_c>0$ and $dq/dr_c<0$ correspond to stable and unstable circular orbits, respectively. 
It is easy to see that the massless curve intersects with the curve~(\ref{eq:ISCO-condition-1}) at $(\qstar,r_\star)$ where we have $dq/dr_c=0$.
The curve~(\ref{eq:massless-Vr-expr}) also intersects with the horizon curve $f=0$ at the point $(q,r)=(q^{\ex},r_h^{\ex})$, since 
\begin{eqnarray}
V_{,r}(q^{\ex},r_h^{\ex})=\frac{rf'(q^\ex,r_h^\ex))-2f(q^{\ex},r_h^{\ex})}{(r_h^{\ex}{})^3}=0.
\end{eqnarray}

\medskip
Therefore, the massless curve has two branches,  $r^+_\infty(q)$ for $r_\star \le r\le 3r_g/2$ and $r^-_\infty(q)$ for $r^\ex \le r\le r_\star$, which correspond to unstable circular orbits and stable circular orbits for massless particles, respectively. 

\subsection{$\mu=3$ case}
In general, the condition~(\ref{eq:massless-Vr-expr}) is difficult to solve explicitly. However, we find that the $\mu=3$  case reduces to the problem finding intersections between the following two graphs $y=F(x)$ and $y=G(x)$,
\begin{eqnarray}
(r_c^\nu+q^\nu)^{\frac{3}{\nu}+1}= \frac{3}{2} r_g \mu r_c^{\nu+2}
\Longleftrightarrow 
\left\{
\begin{array}{ll}
 &y= F(x):=(x+\tilde q)^{\nu+3}\\
 &y=G(x):= (3/2)^\nu   x^{\nu+2}
\end{array}
\right. .
\end{eqnarray}
where $x:=(r_c/r_g)^\nu$ and $\tilde{q}:=(q/r_g)^\nu$.
In particular, the critical value $\qstar$ and $r_\star$ is given by $F(x)=G(x)$ and $F'(x)=G'(x)$, which are analytically solved as
\begin{eqnarray}
r_\star/r_g=\frac{3}{2} \left(\frac{\nu+2}{\nu+3}\right)^{\frac{\nu+3}{\nu}},\quad 
\qstar/r_g = \frac{3}{2}\frac{(\nu+2)^{\frac{\nu+2}{\nu}}   }{(\nu+3)^{\frac{\nu+3}{\nu}}     }.
\end{eqnarray}

\section{Circular photon orbits}\label{sec:photon}
Under the NED Lagrangian $\cL(\cF)$, photons do not propagate along null geodesics of the spacetime geometry, but rather of the so-called effective geometry~\cite{Novello:1999pg}.
The eikonal limit for photons leads to the condition
\begin{align}
 \tilde{g}_{\mu\nu} k^\mu k^\nu =0,
\end{align}
and $\tilde{g}_{\mu\nu}$ is the effective geometry given by
\begin{align}
\tilde{g}^{\mu\nu} = g^{\mu\nu} - \frac{4\cL_{\cF\cF}}{\cL_\cF} F^\mu{}_\alpha F^{\alpha \nu},\label{eq:eff-geom}
\end{align}
where ${\cal L}_{\cal F} := d{\cal L}/d{\cal F}$ and ${\cal L}_{\cal F F}:=d^2{\cal L}/d{\cal F}^2$.
In the magnetic case, with $\cL(\cF)$ and $F_{\mu\nu}$ in eqs.~(\ref{eq:LF-magnetic}) and (\ref{eq:F-magnetic}), the effective metric is written as
\begin{align}
\tilde{g}_{\mu\nu}^{(m)} = {\rm diag}\left(-f,\frac{1}{f},\frac{r^2}{\Phi^{(m)}},\frac{r^2\sin^2\theta}{\Phi^{(m)}}\right),
\end{align}
where $\Phi^{(m)}:= 1+2 \cL_{\cF\cF} \cF/\cL_\cF|_{\rm magnetic}$. On the other hand, the electric solution admits the effective metric of
\begin{align}
\tilde{g}_{\mu\nu}^{(e)} = {\rm diag}\left(-\frac{f}{\Phi^{(e)}},\frac{1}{f \Phi^{(e)}},r^2,r^2\sin^2\theta\right),
\end{align}
where $\Phi^{(e)}:=1+2 \cL_{\cF\cF} \cF/\cL_\cF|_{\rm electric}$ is now given by the electric counterparts in eqs.~(\ref{eq:A-electric}) and (\ref{eq:L-electric}). It is known that these two cases cannot be distinguished by the photon propagation~\cite{Toshmatov:2021fgm}.
This fact can be seen from a certain kind of duality in NED with the same metric,
\begin{align}
\cL_\cF^2 \cF|_{\rm electric} = -\cF|_{\rm magnetic},
\quad \cL_\cF|_{\rm magnetic} = (\cL_\cF)^{-1}|_{\rm electric}.
\end{align}
which leads to
\begin{align}
 \Phi^{(m)} = \frac{1}{\Phi^{(e)}},
\end{align}
Therefore, the effective geometries for the electric and magnetic solutions
are related through the conformal transformation\footnote{We do not consider the point $\Phi^{(m)}=0$, where the effective geometry is singular~\cite{Novello:2000km}. The eikonal limit will not be appropriate there.}
\begin{align}
 \tilde{g}_{\mu\nu}^{(e)} = \Phi^{(m)}\tilde{g}_{\mu\nu}^{(m)}.
\end{align}
and hence have the same causal structure.
Thus, the effective metrics for both the electrically and magnetically charged spacetimes are different, 
but the effective potentials for photons can be shown to be the same, i.e.,
\begin{align}
& \tilde{V} = \frac{f}{r^2 \Phi^{(e)}}= \frac{f}{r^2} \Phi^{(m)} .
\end{align}
For this reason, the photon trajectories coincide in both effective geometries.

With eqs.~(\ref{eq:LF-magnetic}) and (\ref{eq:F-magnetic}), the effective potential becomes
\begin{align}
\tilde{V} = \frac{f}{r^2}\left[\frac{\left(\mu ^2-4 \mu +3\right) q^{2 \nu }-\left(\mu  (3 \nu +4)+\nu ^2-4 \nu -6\right) q^{\nu } r^{\nu }+\left(\nu ^2+4 \nu +3\right) r^{2 \nu }}{2 \left(q^{\nu }+r^{\nu
   }\right) \left((\nu +3) r^{\nu }-(\mu -3) q^{\nu }\right)}\right].
\end{align}

\begin{figure}
\begin{center}
\includegraphics[width=8cm]{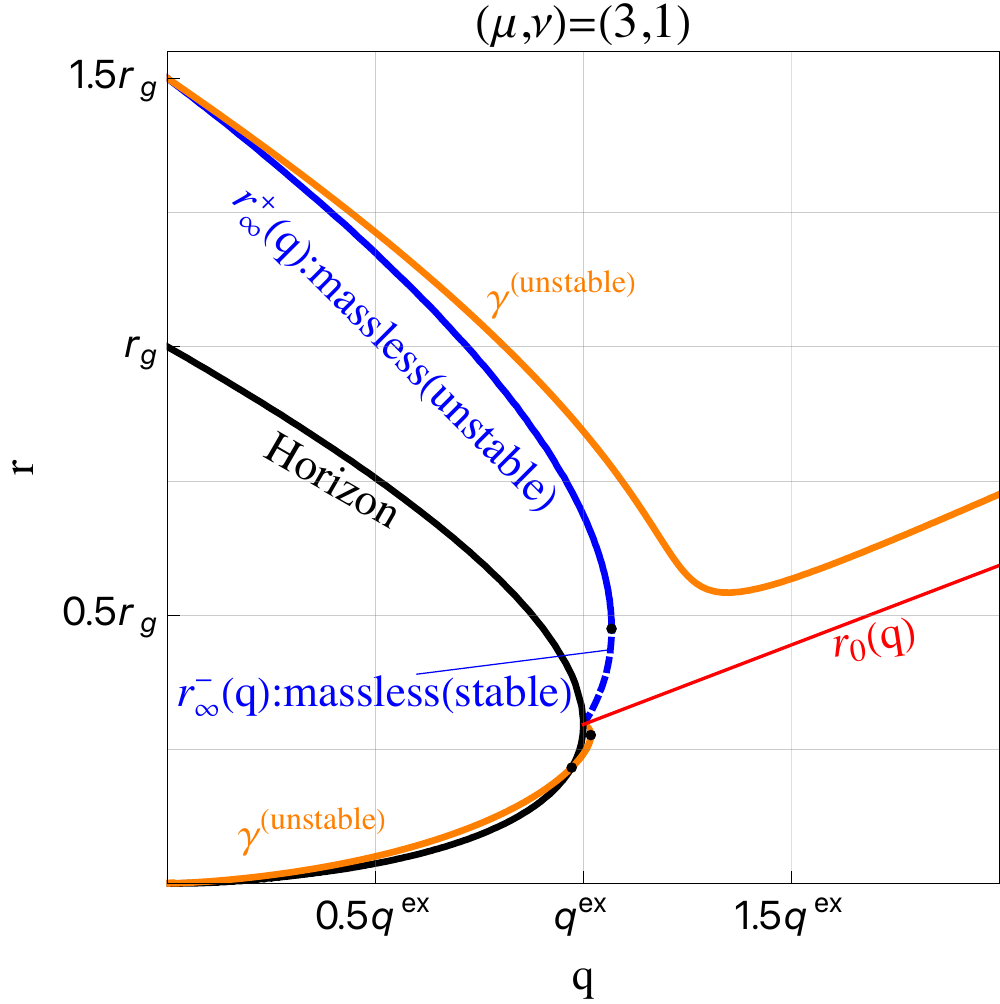}\hspace{0.5cm}
\includegraphics[width=7cm]{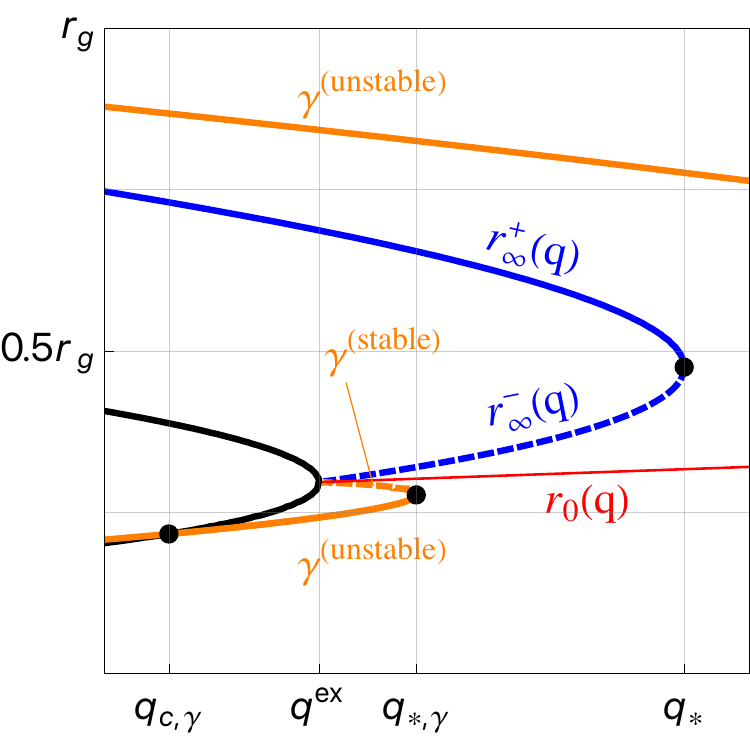}
\caption{Circular photon orbits and massless orbits for $(\mu,\nu)=(3,1)$.  The right figure is a closeup around $q=q^\ex$. The $L=0$ line for massive particles is also drawn for reference. \label{fig:photonorbits}}
\end{center}
\end{figure}

\begin{figure}
\begin{center}
\includegraphics[width=7.8cm]{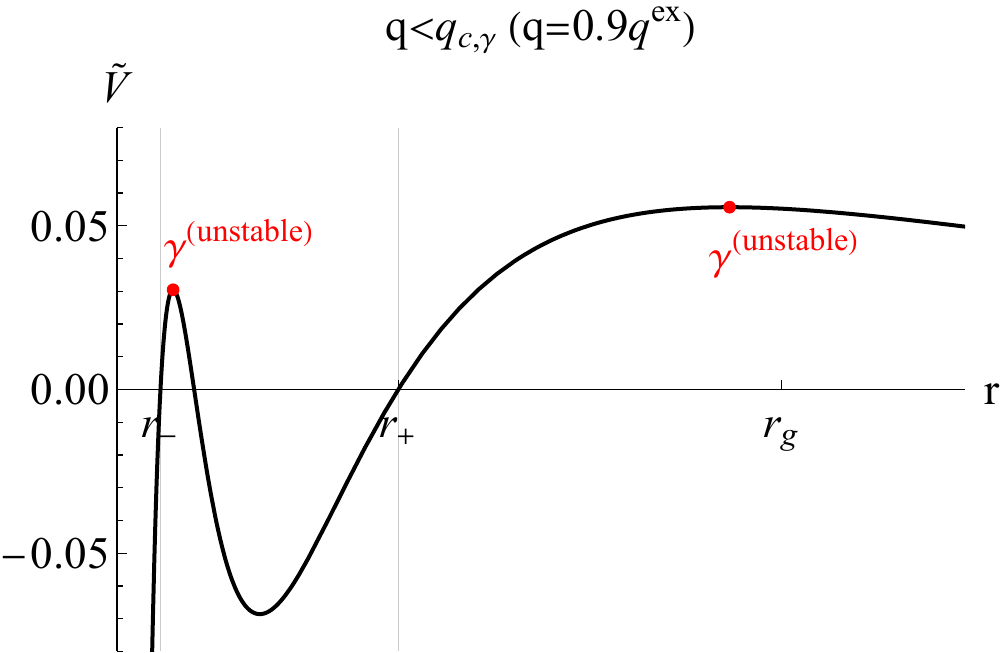}\quad
\includegraphics[width=7.8cm]{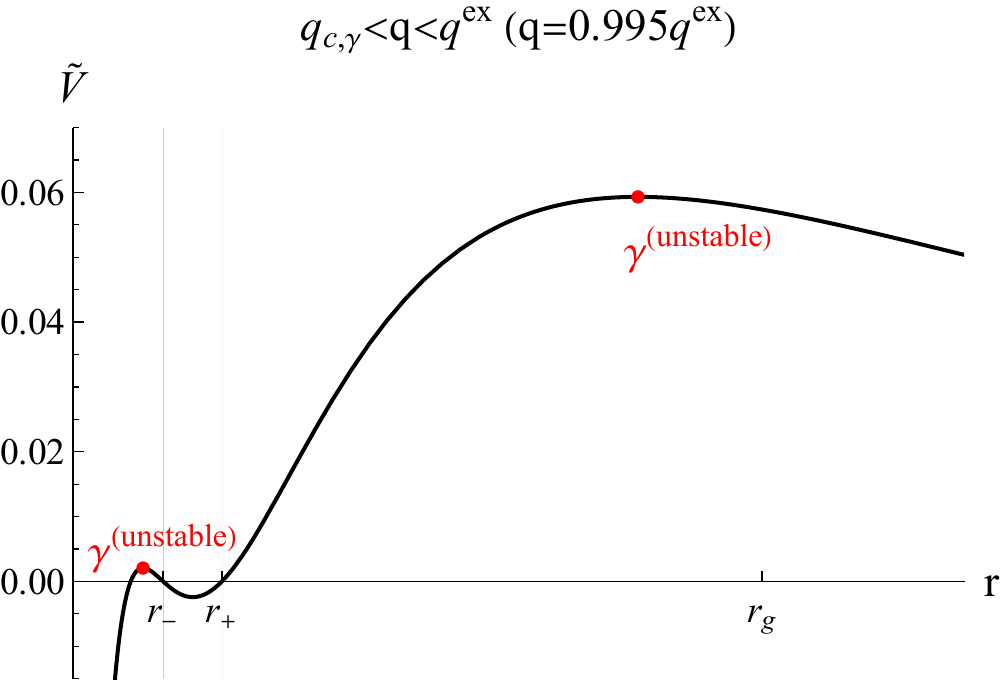}\\
\includegraphics[width=7.8cm]{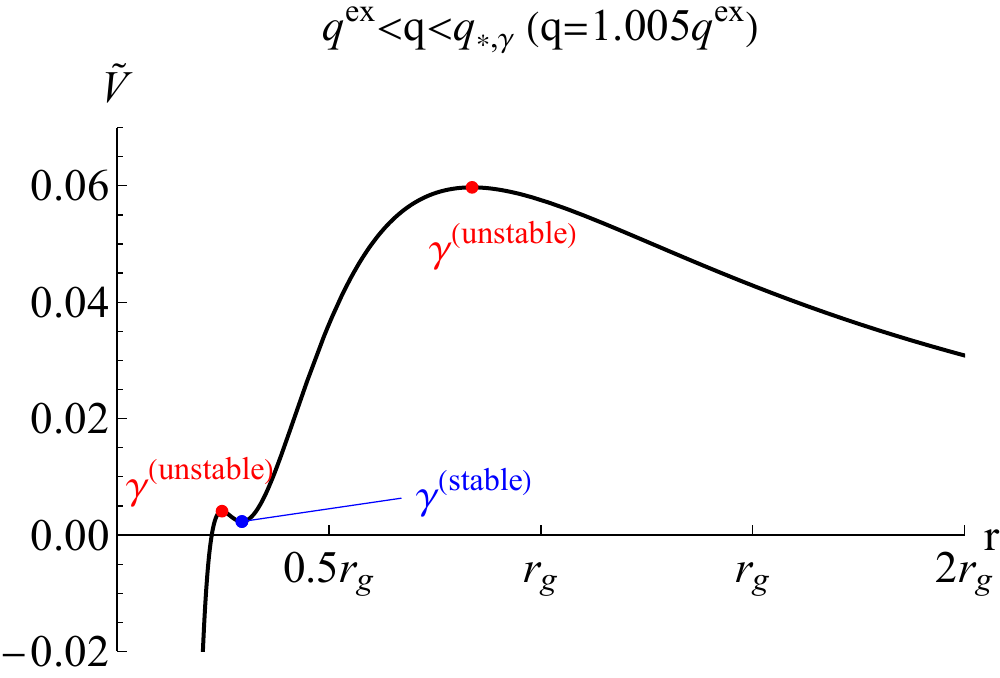}\quad
\includegraphics[width=7.8cm]{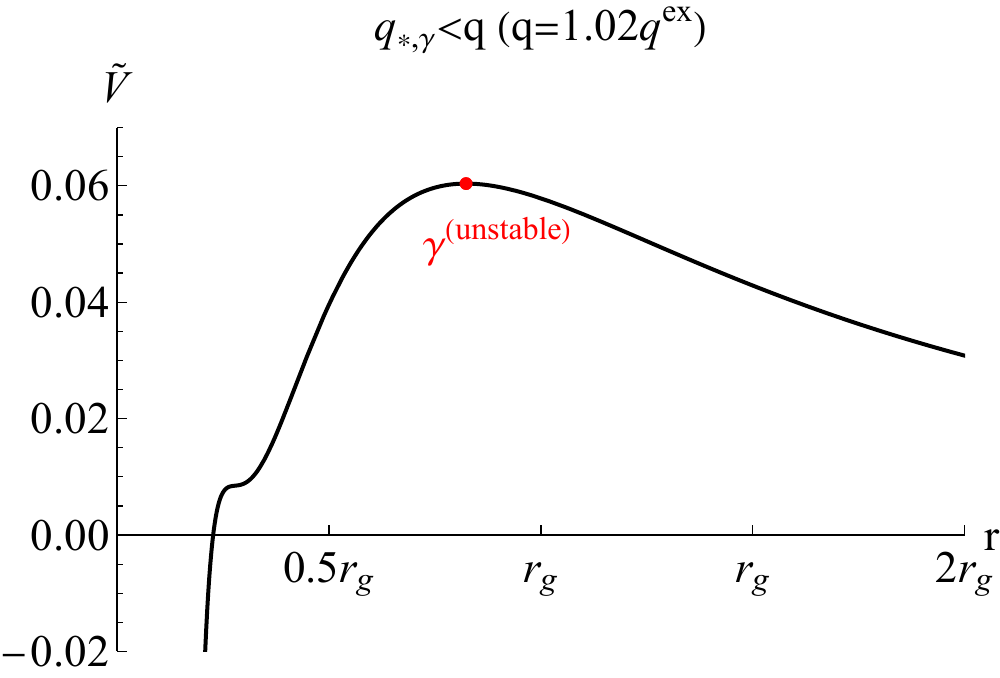}
\caption{The effective potential for photon orbits with $(\mu,\nu)=(3,1)$ and different charges.\label{fig:Vphotons}}
\end{center}
\end{figure}

\begin{figure}
\begin{center}
\includegraphics[width=16cm]{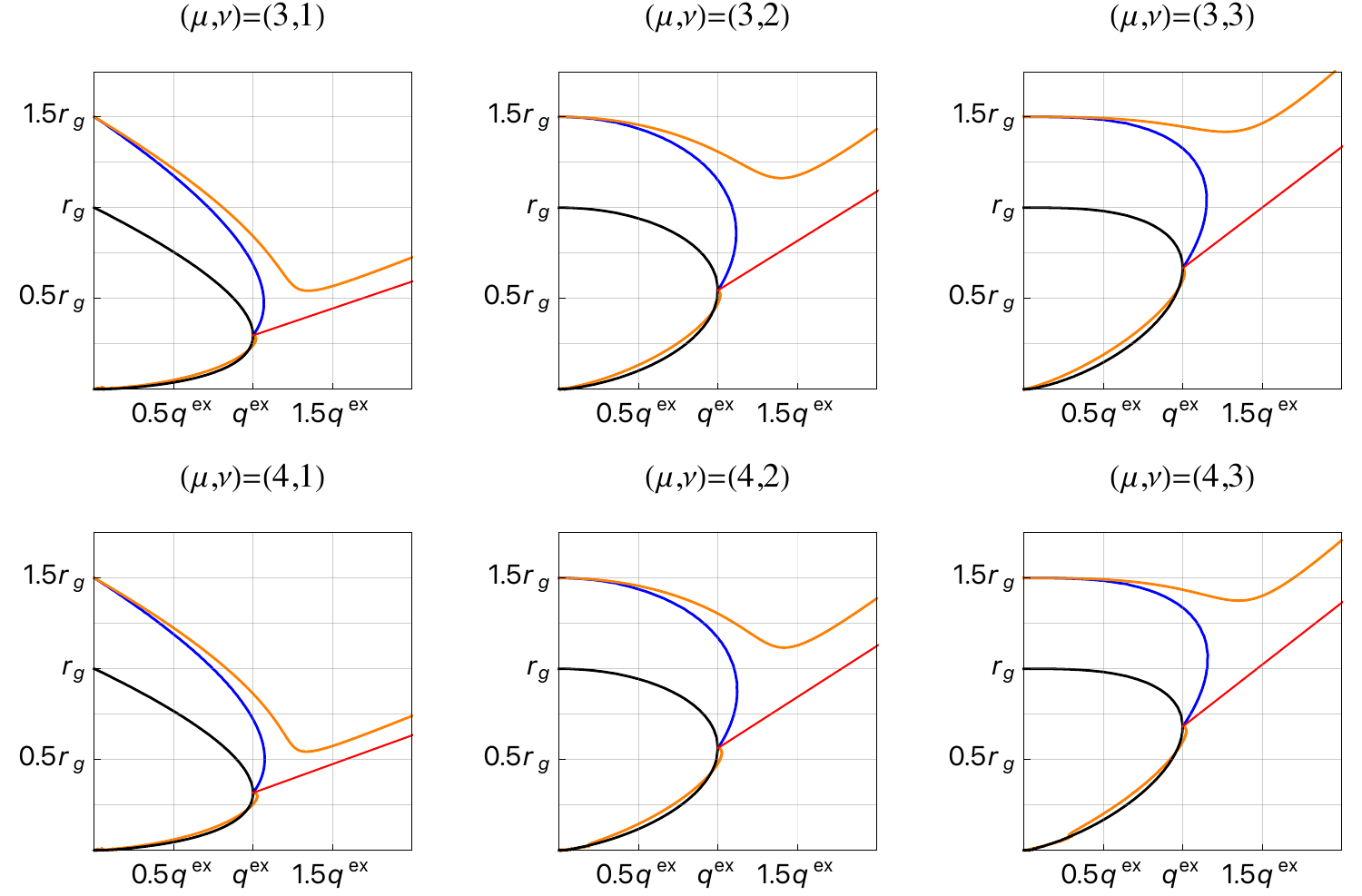}
\caption{Comparison between photon orbits and massless orbits. The convention is the same as in Fig.~\ref{fig:photonorbits}. \label{fig:photonorbitsmore}}
\end{center}
\end{figure}

In Fig.\ref{fig:photonorbits}, we compare the circular photon orbits with that of massless particles for $(\mu,\nu)=(3,1)$.
We find an unstable photon orbit always exists for any $q$ outside the massless orbit as seen in the $\nu=1$ case~\cite{Stuchlik:2019uvf}.
For $q>q^\ex$, this unstable orbit still appears outside the ISCO radius $r_0(q)$.
Below a certain critical charge $q < \qstar_{,\gamma}$, we also find another two orbits inside the ISCO radius $r_0(q)$. The upper stable branch only exists in the overcharged case $q^\ex<q<\qstar_{,\gamma}$, while the lower unstable branch exists with and without the horizon. The lower branch crosses the horizon at $(q_{c,\gamma},r_{c,\gamma})$. 
Remarkably, for $0<q<q_{c,\gamma}$, the lower branch gives a circular orbit between the inner and outer horizon, where no stationary motion is allowed in the spacetime geometry. For $q_{c,\gamma}<q<q^\ex$, the orbit appears inside the inner horizon.
In Fig.~\ref{fig:Vphotons}, the typical shapes of the effective potential are shown corresponding to the range of $q$.
We also obtain qualitatively the same results for other parameters~(Fig.~\ref{fig:photonorbitsmore}).

\section{Periapsis shifts}\label{sec:shift}
Next, we consider the perihelion shift in the massive particle orbits. In ref.~\cite{Gao:2020wjz}, the shift was numerically studied for the orbits close to the horizon. In this article, instead, we focus on the distant orbits where the weak field limit is available to find the analytic formula for the shift.

First, we expand the effective potential~(\ref{eq:geodesic-r-U}) at the large distance from the horizon by assuming $r \gg r_+ \sim r_g$ and $q/r\ll1$,

\begin{align}
& U = 1 - \frac{r_g}{r} + \frac{L^2+r_g \mu q}{r^2} -\left(r_g L^2 + \frac{r_g \mu (\mu-1)q^2}{2}\right)\frac{1}{r^3}+{\cal O}\left(\frac{1}{r^{4}}\right)\quad &(\nu=1),\label{eq:Uexpand-nu1}\\
& U = 1 - \frac{r_g}{r} + \frac{L^2}{r^2} -\left(r_g L^2 - \frac{r_g \mu q^2}{2}\right)\frac{1}{r^3}+{\cal O}\left(\frac{1}{r^{4}}\right)\quad& (\nu=2),\\
& U = 1 - \frac{r_g}{r} + \frac{L^2}{r^2} -\frac{r_g L^2}{r^3}+{\cal O}\left(\frac{1}{r^{4}}\right) \quad &(\nu \geq 3),
\end{align}
where the dominant terms depends on the parameter $\nu$.
Since the leading order correction coincides with that of the Schwarzschild, we will not consider $\nu \geq 3$ cases. In the following, we study $\nu=1$ and $\nu=2$ cases.
Note that, for the valid expansion, $q$ should not be much large
\begin{align}
 \frac{L^2}{r_g^2} \gg \frac{\mu q}{r_g}.\label{eq:q-range-PN}
\end{align}

\subsection{$\nu=1$}
In the $\nu=1$  case, the effective potential already differs in the Newtonian order,
\begin{align}
 U = 1 - \frac{r_g}{r} + \frac{\alpha^2 \ell^2 r_g^2}{r^2}+{\cal O}\left(\frac{1}{r^3}\right),
\end{align}
where we introduced dimensionless parameters
\begin{align}
\alpha := \sqrt{1 + \frac{\mu r_g q}{L^2}},\quad \ell := \frac{L}{r_g}.
\end{align}
Therefore, the orbit at the Newtonian order can be solved as
\begin{align}
r = \frac{2 \alpha^2 \ell^2 r_g }{1+  e \cos (\alpha \phi)},\quad e := \sqrt{4\alpha^2\ell^2 (\cE^2-1)+1}
\end{align}
For $\alpha > 1$, this orbit causes the precession in the Kepler motion already at the Newtonian order, which appearantly seems to result in a retrograde shift that is opposite to the known Schwarzschild result
\begin{align}
\delta \phi = \frac{2\pi}{\alpha}-2\pi = 2\pi \left(\frac{1}{\sqrt{1+\frac{\mu q}{r_g \ell^2} }}-1\right)<0.
\end{align}
However, for the BH case $q<q^\ex$, since the weak field limit requires $\ell \gg 1$, $\alpha$ must be close to $1$,
\begin{align}
 \alpha \simeq 1 + \frac{\mu q}{2r_g \ell^2},
\end{align}
which leads to the expression
\begin{align}
 \delta \phi \simeq - \frac{\pi \mu q}{ r_g \ell^2}.
\end{align}
Note that the this approximation is independent on the value of $\mu$ from eq.~(\ref{eq:qex-range}). This is in the order of $\ell^{-2}$ that is the same order of the post-Newtonian correction from the $L^2/r^3$ term. Hence, we have to take $L^2/r^{3}$ correction into account as done in the Reissner-Nordstr\"om spacetime~\cite{Hong:2017dnf},
\begin{align}
 \delta \phi \simeq  - \frac{\pi \mu q}{ r_g \ell^2} + \frac{3\pi }{2\ell^2}= \frac{\pi}{\ell^2} \left(\frac{3}{2} - \frac{\mu q}{r_g}\right)
\end{align}
where we ignored the second term in the coefficient of $r^{-3}$ in eq.~(\ref{eq:Uexpand-nu1}) as it becomes of ${\cal O}(\ell^{-4})$.
Using eq.~(\ref{eq:qex-range}), we obtain the upper bound for the charge term
\begin{align}
\mu q/r_g \leq \mu  q^\ex/r_g \leq 4/9,
\end{align}
which indicates the shift remains prograde, $\delta \phi \geq 19\pi/(18\ell^2)$, even at the extremal limit.

If we consider the overcharged case $q>q^\ex$, we have no upper bound for $q$. Therefore, the shift can become retrograde for $q > 3r_g/(2 \mu)$.\footnote{If one considers $q$ is simply a cut off scale of the quantum gravity origin, it should be in the Planckian order and it is unphysical to discuss the overcharged case. Here we consider $q$ simply as the charge in NED.}

\subsection{$\nu=2$}
In the $\nu=2$ case, the difference only appears in the post-Newtonian correction order
\begin{align}
 U = 1-\frac{r_g}{r} + \frac{\ell^2 r_g^2}{r^2} - \frac{(\ell^2-\ell_c^2) r_g^3}{r^3}+{\cal O}\left(\frac{1}{r^4}\right),
\end{align}
where we introduced a dimensionless parameter
\begin{align}
 \ell_c :=  \sqrt{\frac{\mu}{2}} \frac{q}{r_g}.
\end{align}
This shows the charge $q$ slightly lowers the shift as in the $\nu=1$ case,
\begin{align}
 \delta \phi = \frac{3\pi}{2\ell^2}\left(1-\frac{\ell^2_c}{\ell^2}\right).\label{eq:dphi-nu2}
\end{align}
A caveat is that, eq.~(\ref{eq:qex-range}) shows $\ell_c$ is bounded above in the BH case as $\ell_c \leq\sqrt{2}/3$, and then eq.~(\ref{eq:q-range-PN}) requires $\ell \gg \ell_c$ even for the overcharged case. Although it does not change the conclusion on the charge effect, this implies the charge correction is of ${\cal O}(\ell^{-4})$, and then one should add the correction from next post-Newtonian order in eq.~(\ref{eq:dphi-nu2}) for the correct estimate.

\section{Summary}\label{sec:conclusion}
In this article, we have investigated the geodesic motion of massive/massless particles and photons around general Fan-Wang spacetimes. For massive and massless particles, we have found that the characteristics of the motions are classified into four cases depending on the strength of the charge (i) $0\leq q\leq q_{\ex}$, (ii)$q_{\ex} < q < \qstar$, (iii)$q_\star \leq q < \qdstar$, (iv)$\qdstar \leq q$. The case (i) has the ISCO outside the horizon, while the cases (ii)-(iv), which is horizonless, have the ISCO with zero angular momentum where the repulsion from the de Sitter core and the gravitational attraction balance.

\medskip
The circular photon orbits are also studied
by examining the null geodesics in the effective geometry.
We found three types of orbits, outer, middle and inner orbits. 
The outer orbit is unstable and always exists for any charge $q$.
The middle one is stable and only exist in the range $q^\ex < q < q_{\star,\gamma}$. The inner one is unstable joining with the middle one at $q_{\star,\gamma}$ and exists for $0<q\leq q_{\star,\gamma}$ with and without the horizon. Remarkably, the inner unstable orbit appears between the inner and outer horizon for $0<q<q_{c,\gamma}$, where no stationary motion is allowed in the spacetime geometry. 

We have also studied the periapsis shift by the massive particle.
We have found the shift is characterized by the parameter $\nu$ as
\begin{enumerate}
\item $\nu=1$ : the shift gets the negative correction, which can change the sign of the shift for the overcharged case
\item $\nu=2$ : the shift gets the negative correction, which remains small in the weak field limit
\item $\nu\geq 3$ : the charge effect is ignorable compared to the GR effect
\end{enumerate}

\medskip
We found that the massless particles and photons can move along stable circular orbits for slightly overcharged spacetimes.
Since the existence of stable null circular orbits is known to cause an instability in the spacetime~\cite{Keir:2014oka,Cardoso:2014sna,Cunha:2017qtt}, 
it would be interesting to pursue the final state of the spacetime with such orbits.

\medskip
The optics inside the horizon would be another interesting subject, due to the existence of the circular orbits of photons.
Other than circular orbits, one can also study the motion of particles and photons falling into the event horizon, which may be an interesting issue as well.

\acknowledgments
The authors thank Tomohiro Harada, Ken-ichi Nakao and Hideki Maeda for useful comments and discussion. The authors also thank Daniele Malafarina
for providing useful comments on the photon orbit.
This work is supported by Toyota Technological Institute Fund for Research Promotion A. 
RS was supported by JSPS KAKENHI Grant Number JP18K13541. 
ST was supported by JSPS KAKENHI Grant Number 21K03560.




\end{document}